\documentclass[twocolumn]{aastex62}
\usepackage{multirow}
\usepackage{makecell}

\def\lax{{$\mathrel{\hbox{\rlap{\hbox{\lower4pt\hbox{$\sim$}}}\hbox{$<$}}}$}}
\def\gax{{$\mathrel{\hbox{\rlap{\hbox{\lower4pt\hbox{$\sim$}}}\hbox{$>$}}}$}}

\begin{document}

\title{On the Connection Between Spiral Arm Pitch Angle and Galaxy Properties}

\author[0000-0002-3462-4175]{Si-Yue Yu}
\affiliation{Kavli Institute for Astronomy and Astrophysics, Peking University, Beijing 100871, China}
\affiliation{Department of Astronomy, School of Physics, Peking University, Beijing 100871, China}

\author[0000-0001-6947-5846]{Luis C. Ho}
\affiliation{Kavli Institute for Astronomy and Astrophysics, Peking University, Beijing 100871, China}
\affiliation{Department of Astronomy, School of Physics, Peking University, Beijing 100871, China}

\begin{abstract}
We measure the pitch angle ($\varphi$) of spiral arms in a sample of 79 
galaxies to perform a systematic study of the dependence of $\varphi$ on 
galaxy morphology, mass, and kinematics to investigate the physical origin of 
spiral arms.  We find that $\varphi$ decreases (arms are more tightly wound), 
albeit with significant scatter, in galaxies with earlier Hubble type, more 
prominent bulges, higher concentration, and larger total galaxy stellar mass 
($M_*^{\rm gal}$).
For a given concentration, galaxies with larger stellar masses 
tend to have tighter spiral arms, and vice versa.  We also find that $\varphi$ 
obeys a tight inverse correlation with central stellar velocity dispersion for
$\sigma_c\ga100$\,km\,s$^{-1}$, whereas $\varphi$ remains approximately 
constant for $\sigma_c\lesssim100$\,km\,s$^{-1}$.  We demonstrate that 
the $\varphi$\,-\,$\sigma_c$ and $\varphi$\,-\,$M_*^{\rm gal}$ relations 
are projections of a more fundamental three-dimensional 
$\varphi\,-\,\sigma_c\,-\,M_*^{\rm gal}$ relation, such that pitch angle 
is determined by $\sigma_c$ for massive galaxies but by $M_*^{\rm gal}$ 
for less massive galaxies. Contrary to previous studies, we find that $\varphi$
correlates only loosely with the galaxy's shear rate.  For a given shear rate, 
spirals generated from $N$-body simulations exhibit much higher $\varphi$ than 
observed, suggesting that galactic disks are dynamically cooler (Toomre's $Q 
\approx 1.2$).  
Instead, the measured pitch angles show a much stronger relation with morphology 
of the rotation curve of the central region, such that galaxies with centrally peaked rotation curves
have tight arms, while those with slow-rising rotation curves have looser arms.
These behaviors are qualitatively consistent with predictions of density wave theory.

\end{abstract}


\section{Introduction}

Spiral structures are the most prominent features of disk galaxies, but their 
physical origin is still debated.  Pitch angle ($\varphi$), defined as the 
angle between the tangent of the spiral and the azimuthal direction, describes
the degree of tightness of a spiral arm \citep{Binney}.  

As the tightness of spiral arms constitutes one of the essential criteria of 
Hubble's scheme of morphological classification of galaxies \citep{Hubble1926, 
Sandage1961}, the pitch angle of spiral arms tends to decrease (become more 
tightly wound) from late to early Hubble types \citep{Kennicutt1981, Ma2002, 
Yu2018a}. Density wave theory \citep{LinShu64, Bertin1989, Bertin1989b, 
Bertin1996} offers the most successful framework to explain spiral structures. 
The semi-empirical study of \cite{Roberts1975} showed that density wave theory 
can fit observed spiral arms and suggested that mass concentration is the main 
determinant of spiral arm pitch angle.
By constructing appropriate basic states of galaxy models, \cite{Bertin1989} 
numerically calculated density wave modes that are able to represent all 
Hubble types and confirmed that spiral arms become tighter with increasing 
mass fraction in the central spherical component. 
A number of studies have used other lines of evidence 
to argue in favor of density wave theory, in terms of the classic aging of 
the stellar population that produces a color gradient across spiral arms 
\citep{Gonzalez1996, Martinez2009a, Martinez2011, Yu2018b} and the dependence 
of pitch angle on wavelength predicted by \cite{Gittins2004}
\citep{Martinez2012,Martinez2013,Martinez2014,Yu2018b}.

In contrast, $N$-body simulations of isolated pure stellar disk galaxies 
generate spiral arms as transient but recurrent structures \citep{Carlberg1985,
Bottema2003, Sellwood2011, Fujii2011, Grand2012a, Grand2012b, Baba2013, 
Onghia2013}.  In these simulations, the tendency for the number of arms to 
increase with decreasing disk mass fraction and for the pitch angle to increase
with decreasing velocity shear are roughly consistent with the predictions of 
swing amplification theory \citep{Julian1966, Goldreich1978, Toomre1981}.  In 
this picture, wherein spirals are transient, the pitch angle reflects
the effects of differential rotation alone.

Within this backdrop, it is instructive to elucidate in a quantitative manner 
how spiral arm pitch angle relates to various galaxy properties.  Many attempts
have been made, with mixed results.  Pitch angles of spiral arms were found to 
correlate strongly with maximum rotation velocity \citep{Kennicutt1981} and 
mean rotation velocity over the region containing spiral arms 
\citep{Kennicutt1982}, with larger rotational velocities leading to more 
tightly wound arms.   The more recent analysis of \cite{Kendall2015}, however, 
does not support this, although their sample contains only 13 objects. 
Similarly, \cite{Kennicutt1981} found that pitch angle decreases with brighter 
absolute magnitude. 
\cite{Kennicutt1981} further noted a correlation between pitch angle and 
Morgan's (\citeyear{Morgan1958}, \citeyear{Morgan1959}) classification, which
is primarily based on subjective estimates of bulge-to-disk ratio, such that 
galaxies with larger bulge fraction tending to have smaller pitch angle, but 
with considerable scatter.
Paradoxically, this trend is not corroborated when concentration index is used 
as a proxy of concentration of mass \citep{Seigar1998, Kendall2015}.
\cite{Seigar2005, Seigar2006} reported a tight connection between pitch angle 
and morphology of the galactic rotation curve, quantified by the shear rate, 
with open arms associated with rising rotation curves and tightly wound arms 
connected to flat and falling rotation curves.   On the other hand,
\cite{Kendall2015} disputed the tightness of the correlation.  Furthermore,
\cite{Yu2018a} show that about 1/3 of the pitch angles in \cite{Seigar2006}
have been severely overestimated; the remeasured pitch angles correlate with 
shear rate weakly at best.  The pitch angle of spiral arms is also found to be 
correlated strongly with the galaxy's central stellar velocity dispersion, and 
hence with black hole mass \citep{Seigar2008}, by virtue of the well-known 
relation between black hole mass and bulge stellar velocity dispersion 
(see \citealt{KH13}, and references therein). 
Lastly, to round off this list of confusing and often contradictory results, 
\cite{Hart2017} analyzed a large sample of galaxies selected from the Sloan 
Digital Sky Survey \citep[SDSS;][]{York2000} and found very weak correlations 
between pitch angle and galaxy mass, but the surprising trend that pitch angle 
increases with increasing bulge-to-total mass ratio.  

\begin{figure*}
\figurenum{1}
\centering
\includegraphics[width=14cm]{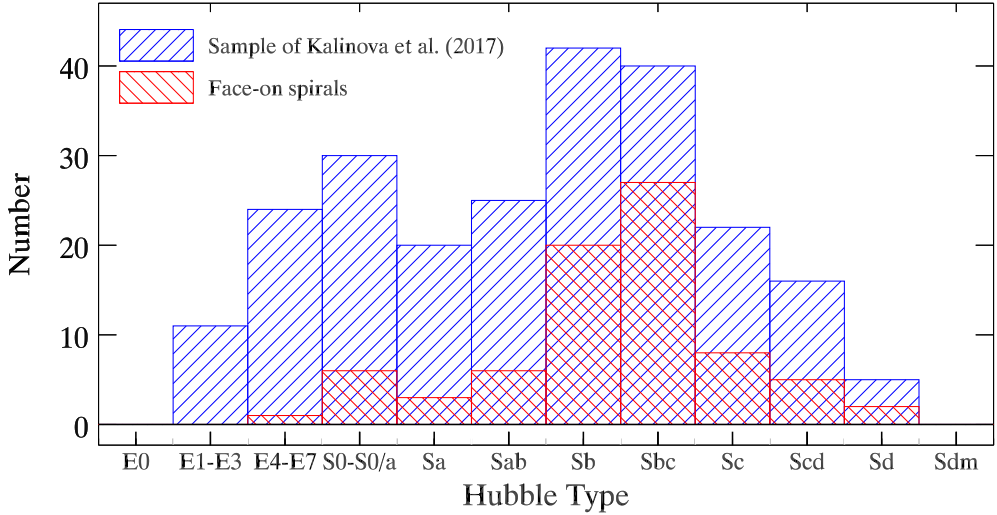}
\caption{Distribution of Hubble types for our sample of 79 galaxies (red-hatched histograms) compared with the sample of 238 objects in \cite{Kalinova2017} (blue-hatched histograms). Our sample spans the full range of Hubble types of disk galaxies, even including an elliptical, which actually has weak but detectable spiral arms.}
\end{figure*}

The morphology of spiral arms may depend on wavelength.  Weak, two-arm spirals 
had been seen in some flocculent galaxies \citep{Block1994, Thornley1996, 
Thornley1997, Elmegreen1999, Block1999}.  Still, pitch angles measured in the 
near-infrared generally agree well with those measured in the optical 
\citep{Seigar2006, Davis2012}. The recent study by \cite{Yu2018b} report a 
mean difference in pitch angle of only $\sim 0.5\degr$ between spiral arms 
observed in $3.6\micron$ and {\it R}-band images, which lays the foundation 
for our use of SDSS {\it r}-band images to measure pitch angle in this paper.

The Calar Alto Legacy Integral Field Area (CALIFA) survey \citep{Sanchez2012} 
targets a diameter-limited sample of galaxies covering all morphological types,
in the redshift range $0.005$\,$<$\,$z$\,$<$\,$0.03$.  For these low redshifts, 
SDSS provides images of adequate qualiy for measuring pitch angle \citep{Yu2018a}. 
\cite{FBJ2017} extracted stellar kinematic maps of 300 CALIFA galaxies 
using the {\tt PPXF} fitting procedure \citep{Cappellari2004}.  From this 
original sample, after discarding galaxy mergers and cases with uncertain 
dynamical models, \cite{Kalinova2017} derived circular velocity curves for 238 
objects using detailed stellar dynamical Jeans modeling \citep{Cappellari2008}.
Other galaxy properties, such as stellar masses, photometric decomposition, and 
star formation rates are also available \citep{Walcher2014, Sanchez2016Pipe3D, 
Sanchez2017, JMA2017, Catalan-Torrecilla2017, Gilhuly2018}.  This database 
enables us to perform a comprehensive study of the dependence of spiral arm 
pitch angle on various galaxy properties.

\section{Data}

We select our galaxies from the sample of \cite{Kalinova2017}, who provide
rotation curves of 238 CALIFA galaxies, which can be used to derive velocity 
shear rates.  We use their corresponding SDSS {\it r}-band images to analyze 
their spiral arms.  We visually inspect the images to exclude ellipticals,
edge-on disks, and irregular systems, finally settling on 93 nearly face-on 
galaxies having spiral structure.  We then generate a mask to exclude 
foreground stars and produce star-cleaned images following the procedures 
described in \cite{Ho2011}. The background level was determined from 10 
randomly selected empty sky regions and then subtracted from the star-cleaned 
images.  As shown in the next section, we successfully measured pitch angles 
for 79 of the 93 galaxies.  The distribution of Hubble types for the final 
sample of 79 galaxies is shown in Figure 1 (red-hatched histograms), which is 
compared with the sample of 238 galaxies of 
\cite[blue-hatched histograms;][]{Kalinova2017}. The Hubble types come from the
CALIFA team \citep{Walcher2014}. The subsample of 79 galaxies well represents 
the distribution of Hubble types of the parent sample from \cite{Kalinova2017};
it even includes an elliptical galaxy, which actually exhibits faint arms.

The stellar masses of the galaxies ($M_*^{\rm gal}$) are from 
\cite{Sanchez2017}, who analyzed the stellar population using the Pipe2D 
pipeline \citep{Sanchez2016BPipe3D}.  \cite{JMA2017} performed two-dimensional 
multi-component photometric decomposition and characterized the main stellar 
substructures (bulge, bar, and disk), whose mass and star formation history 
were studied by \cite{Catalan-Torrecilla2017}.  From these studies we can 
compile the bulge-to-total light ratio ($B/T$), bulge stellar mass 
($M_*^{\rm bul}$), and disk stellar mass ($M_*^{\rm disk}$).  
The uncertainty of $B/T$ is primarily systematic in origin, driven by model assumptions instead of statistical errors from the fitting.  Following 
\citep{Gao2017}, we assign 10\% fractional uncertainty to $B/T$.
The concentration indices ($C_{28}$), derived from the isophotal analysis of
\cite{Gilhuly2018},
have fractional errors pf $3\%$.
The absolute {\it B}-band magnitudes ($M_B$) come from 
HyperLeda \citep{Paturel2003}.  The central velocity dispersions ($\sigma_c$) 
and their uncertainties are calculated as the mean value and standard 
deviation of the velocity dispersion, provided by \cite{FBJ2017}, within 
3$\arcsec$ of the galaxy center.  
\cite{Kalinova2017} performed principal component analysis of the rotation curves of the CALIFA galaxies and provided the coefficient of the first eigenvector (PC$_1$), which quantitatively describes the shape and amplitude of the rotation curve of the central region.  Section 4 shows that PC$_1$ is useful for our study. Table 1 lists the above-described parameters for our sample.

\begin{figure*}
\centering
\includegraphics[width=17.5cm]{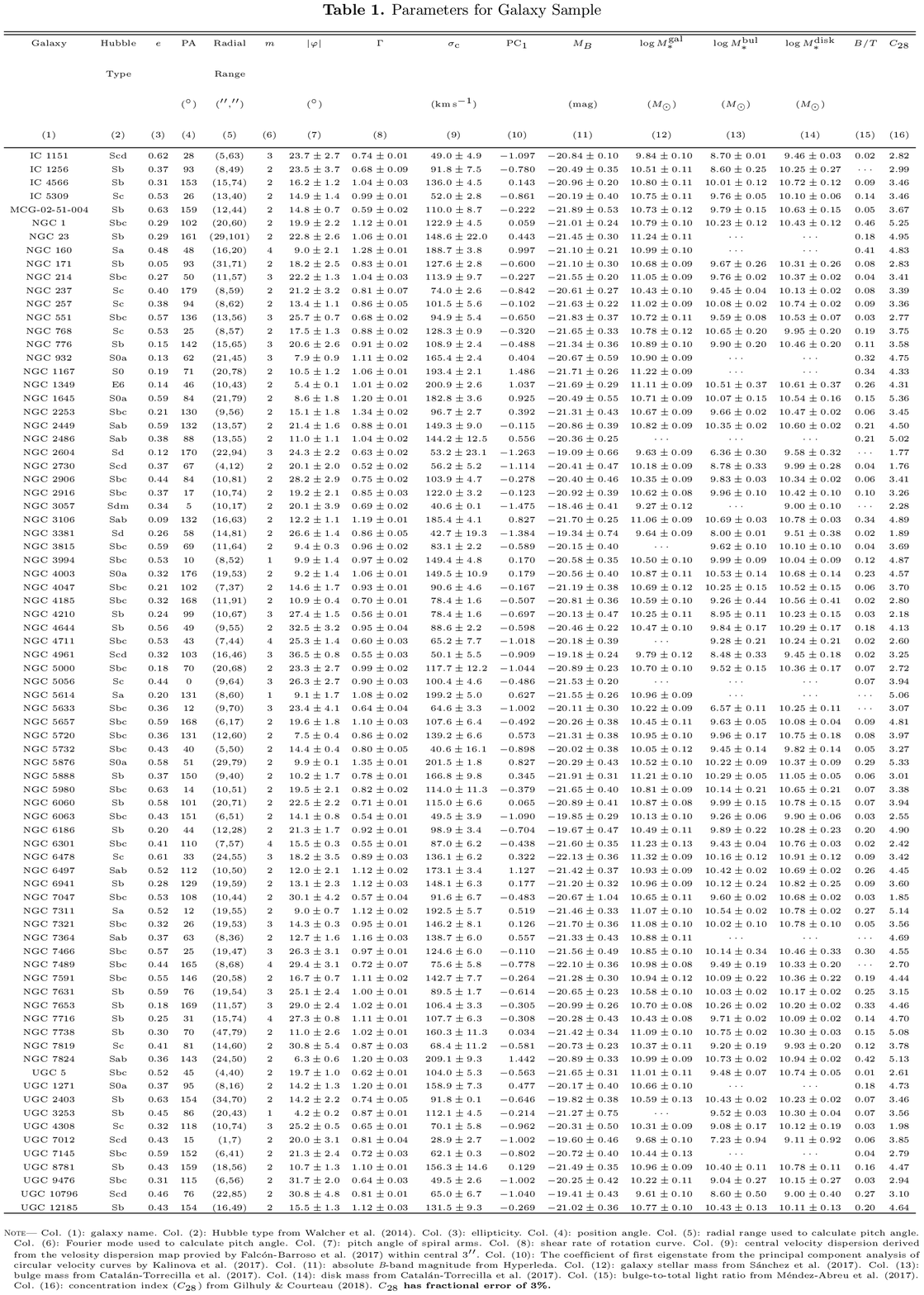}
\end{figure*}

\section{Pitch Angle and Shear Rate}
\subsection{Measuring pitch angle}

An accurate determination of the sky projection parameters---ellipticity ($e$) 
and position angle (PA)--- for the galaxies is essential for the study of 
spiral arms.  We adopt two methods to measure the $e$ and PA of galaxies and 
determine the optimal results. One is to use the {\tt IRAF} task {\tt ellipse} 
to extract radial profiles of $e$ and PA from isophotal analysis. The adopted 
values of $e$ and PA are obtained by averaging their profiles in the region 
where the disk component dominates. The second method is to use a 
two-dimensional Fourier transformation of the disk region, minimizing the real 
part of the Fourier spectra, which corresponds to the bimodal component, to 
derive $e$ and PA. These two methods assume that the disk is intrinsically 
circular. To determine the optimal results, we deproject the galaxies to their
face-on orientation using the $e$ and PA values from these two methods, giving
preference to that which yields a rounder deprojected image or more 
logarithmic-shaped spiral arms. Our adopted values of $e$ and PA are listed in 
Table~1.

The most widely used techniques to measure the pitch angle of spiral arms 
employ discrete Fourier transformation, either in one dimension (1DDFT) 
\citep{Grosbol2004, Kendall2011, Yu2018a} or in two dimensions (2DDFT) 
\citep{Kalnajs1975, Iye1982, Krakow1982, Puerari1992, Puerari1993, Block1999, 
Davis2012, Yu2018a}. \cite{Yu2018a} discuss and use both techniques, in the 
context of images from the Carnegie-Irvine Galaxy Survey \citep[CGS;][]{Ho2011}.
As the pitch angles obtained from both methods are actually consistent within
a small scatter of $2\degr$ \citep{Yu2018a}, we use 2DDFT to measure pitch 
angle for the majority (72/79) of our sample; the 1DDFT method was used for 
seven cases for which the 2DDFT method failed.  In total we successfully 
measured pitch angles for 79 of 93 objects.  Table 1 lists our pitch angle 
measurements, including the radial range and Fourier mode used for the 
calculation.  Pitch angles could not be 
measured for the rest of the galaxies because the arms are too weak (6 
galaxies), too flocculent (4 galaxies), or too wound up such that isophotes
cross a single arm more than once (4 galaxies).  As shown in the distribution 
of Hubble types in Figure 1, $20\%$ (6) of the S0 or S0/a galaxies have very 
weak but detectable spiral arms whose pitch angles can be measured. An extreme
case is NGC~1349, which is classified as ``E'' in \cite{Walcher2014} but ``S0''
in HyperLeda.  Compared with previous studies of spiral arms that use known
``spirals'' as one of their selection criteria \citep{Kennicutt1981, 
Seigar1998, Seigar2005, Seigar2006, Kendall2011, Elmegreen2014}, our sample is 
more complete by including disks in early-type galaxies with faint arms.

\subsection{Measuring shear rate}

The galactic shear rate ($\Gamma$), which quantifies the morphology of the 
rotation curve, is given by

\begin{eqnarray}
\Gamma = \frac{2A}{\Omega} = 2 - \frac{\kappa^2}{2\Omega^2}=1 - (R/V_{c})({\rm d}V_{c}/{\rm d}R),
\end{eqnarray}

\noindent
where $R$ is the radial distance from the center, $V_c$ is the 
circular velocity, $A$ is first Oort constant, $\Omega$ is angular speed, and $\kappa$ is epicyclic frequency. 
Eq. (1) is the most widely adopted definition of shear rate \citep[e.g.,][]{Bertin1989b, 
Grand2013, Dobbs2014, Michikoshi2014},  
but is 2 times the shear rate defined by \cite{Seigar2005}.  We 
use the rotation curves of \cite{Kalinova2017} to derive $\Gamma$.

We first identify an outer region that is beyond the turnover of the rotation 
curve ([$r_{i}$,$r_{o}$]), one that preferably coincides with the radial 
region used to derive the pitch angle. For a few galaxies whose radial region 
occupied by spiral arms exceeds the radial extent of the rotation curve, we 
choose the outer region where the rotation curve has become stable.  To 
evaluate $\Gamma$, we fit the function 

\begin{eqnarray}
V_c=b\times e^{(1-\Gamma)\text{ln}R}, 
\end{eqnarray}

\noindent
where $b$ is a coefficient, to the rotation curve in three radial ranges: 
$[r_i, r_o-\Delta r]$, $[r_i+\Delta r, r_o]$, and $[r_i+\Delta r/2, r_o-
\Delta r/2]$, where $\Delta r=0.2(r_o-r_i)$.  The value of $\Gamma$ and its 
uncertainty (Table 1), introduced from the choice of radial range, are 
estimated as the mean value and standard deviation of the three values of 
$\Gamma$ obtained from the fitting over the above three radial ranges. 

Figure 2 gives an illustration for IC~1511, for which the filled points 
represent the rotation curve and the solid line marks the curve that yields 
a shear rate of $\Gamma=0.74$.  Note that when $\Gamma=1$, the outer part of 
the rotation curve is flat; when $\Gamma<1$, the outer part of the rotation 
curve is rising; when $\Gamma>1$, the outer part of the rotation curve is 
falling.

\begin{figure}
\figurenum{2}
\centering
\includegraphics[width=8cm]{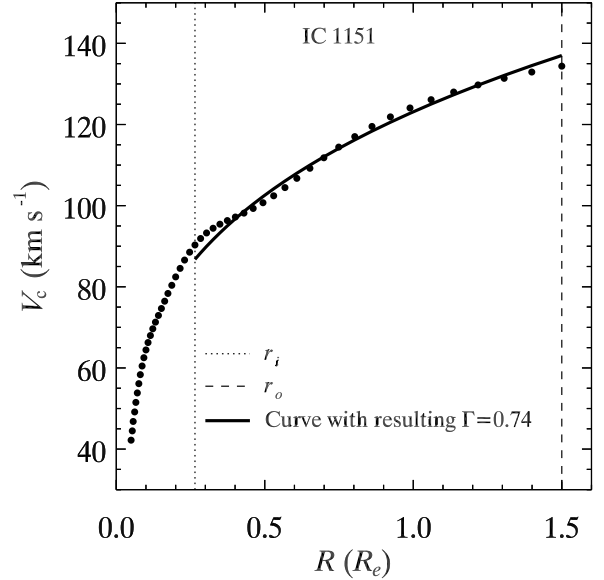}
\caption{Example of measuring shear rate for IC~1151. Three radius ranges: $[r_i, r_o-\Delta r]$, 
$[r_i+\Delta r, r_o]$, and $[r_i+\Delta r/2, r_o-\Delta r/2]$, where $\Delta r=0.2(r_o-r_i)$, are used 
to fit the function in Equation (2). The solid line represents the function with averaged $\Gamma$ and $b$.}
\end{figure}

\section{The Relationship Between Pitch Angle and Galaxy Properties}

In this section, we study the dependence of spiral arm pitch angle on various 
galaxy properties (morphology, luminosity, stellar mass, and kinematics),
and then compare our findings with previous studies, which sometimes reveal 
conflicting results.

\subsection{Dependence on Galaxy Morphology}

\begin{figure}
\figurenum{3}
\centering
\includegraphics[width=7.5cm]{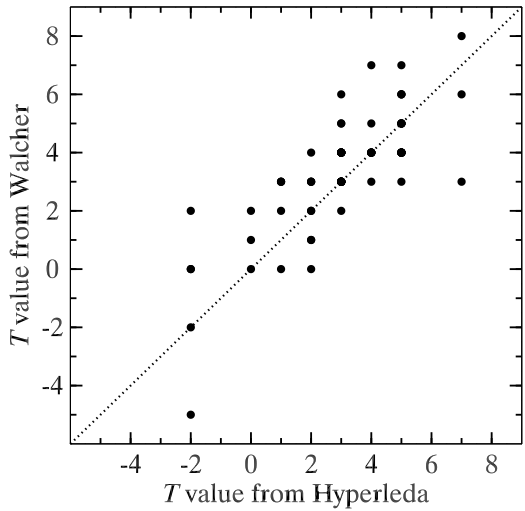}
\caption{ 
Comparison between Hubble type of our sample, adopted from \cite{Walcher2014},
with classifications given in Hyperleda \citep{Paturel2003}.  The correspondence
between Hubble type and $T$ value is as follows: E: $T$\,$=$\,$-5$; S0: 
$T$\,=\,$-2$; S0/a: $T$\,=\,$0$; Sa: $T$\,=\,$1$; Sab: $T$\,=\,$2$; Sb: 
$T$\,=\,$3$; Sbc: $T$\,=\,$4$; Sc: $T$\,=\,$5$; Scd: $T$\,=\,$6$; Sd: 
$T$\,=\,$7$; Sdm: $T$\,=\,$8$; Sm: $T$\,=\,$9$. 
The dotted line mark the $1:1$ ratio.  The Hubble types from these two sources 
are roughly consistent with a total scatter in $T$ value of $\sigma_T = 1.3$.
}
\end{figure}

\begin{figure*}
\figurenum{4}
\centering
\includegraphics[width=16cm]{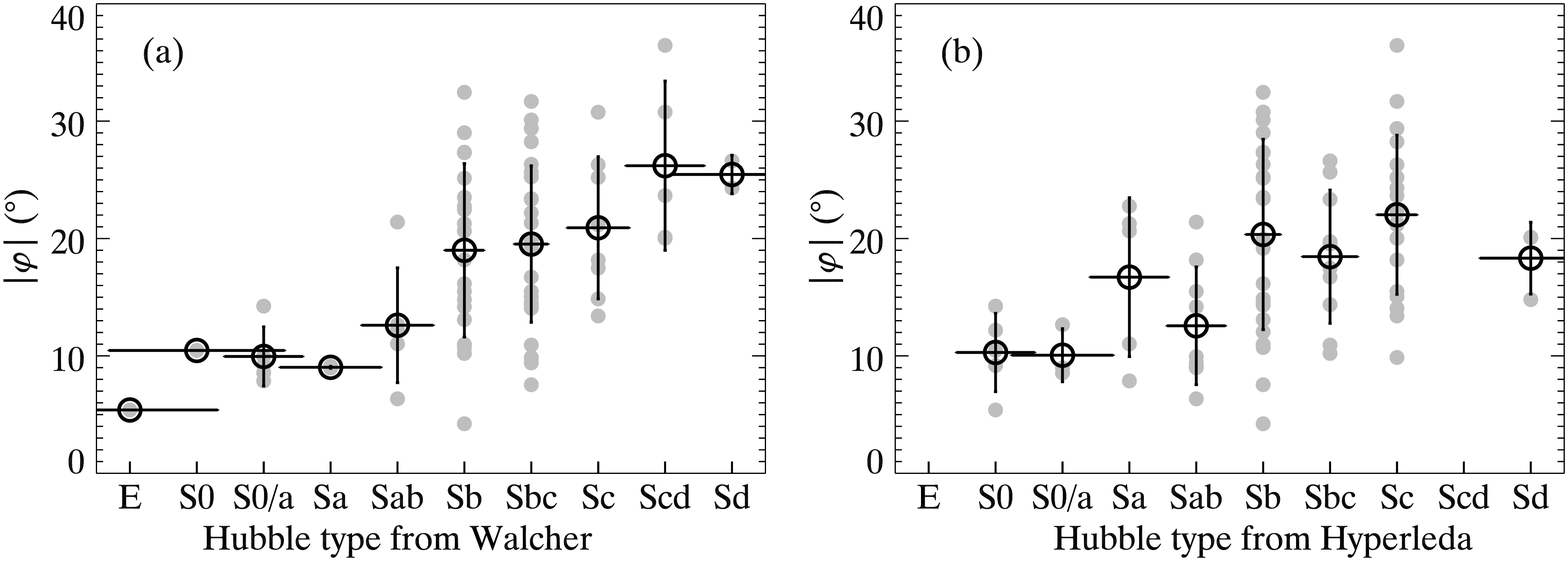}
\caption{Variation of spiral arm pitch angle with Hubble type from (a) \cite{Walcher2014} and (b) Hyperleda \citep{Paturel2003}.  
The large open points mark the mean value and associated errors.  The
uncertainty of the mean Hubble type is determined by $\sigma_T/\sqrt{N}$,  with
$N$ the number of objects in each Hubble type bin.
Most of the scatter in pitch angle for a given Hubble type
is real and not caused by subjective classification.}
\end{figure*}

Hubble types are subjective. Different classifications place different 
weights on the classification criteria and may lead to different results. To 
assess the uncertainty of the Hubble types of our sample, we compare the types 
determined by \cite{Walcher2014} with those given in Hyperleda 
\citep{Paturel2003}: the two are consistent within a scatter of $\sigma_T = 
1.3$ in $T$ (Figure 3).  The measured pitch angle of spiral arms is plotted 
against Hubble type determined by \cite{Walcher2014} in Figure~4a, where the 
open points mark the mean value and associated errors.  The
uncertainty of the mean Hubble type is determined by $\sigma_T/\sqrt{N}$,  with
$N$ the number of objects in each Hubble type bin. 
Despite the fact that our sample contains fewer 
early-type spirals than late-type spirals, our results confirm that on 
average spiral arms tend to be more tightly wound in galaxies of earlier Hubble
type, but with large scatter in pitch angle ($\sim7\degr$).  Most of the 
early-type spirals (Sab and earlier) have pitch angles less than $15\degr$, 
while later type spirals (Sb and later) can have both high ($\sim 30\degr$) 
and low ($\sim 10\degr$) pitch angles.  This behavior is not entirely 
consistent with the Hubble classification system, which implicitly considers 
tightness of spiral arms.  Part of the scatter in the relation between pitch 
angle and Hubble type may come from the subjective nature of morphological 
classification. 
However, repeating
the analysis using Hubble types from Hyperleda \citep{Paturel2003} yields 
very similar results (Figure~4b), indicating that a large portion of scatter 
in pitch angle at a given Hubble type is real.  Studies by \cite{Kennicutt1981}
and \cite{Ma2002} also found a weak correlation between pitch angle and Hubble 
type, with large scatter in pitch angle at a given type.  But such a 
trend was not seen by \cite{Kendall2015}, probably because of their small 
sample of just 13 objects, nor by \cite{Seigar1998}, probably because nearly 
all of their pitch angles were less than $15\degr$.

Density wave theory predicts an inverse correlation between pitch angle and 
mass concentration \citep{LinShu64, Roberts1975, Bertin1989, Bertin1989b}.  The
relative prominence of the bulge, as reflected, for instance, in the 
bulge-to-total light ratio ($B/T$), should provide a reasonable proxy for the
stellar mass concentration.  
The same holds for the concentration parameter $C_{28} \equiv R_{80}/R_{20}$, 
with $R_{20}$ and $R_{80}$ the radii enclosing, respectively, 20\% and 80\% of 
the total flux.   Figure~5 shows the relation between pitch angle $\varphi$ 
and $B/T$ and $C_{28}$.  We group the data points into five equal-sized bins of 
$B/T$ and $C_{28}$, and then calculate the mean value and standard deviation 
of each bin.  
The number of bins is set to ensure sufficient sampling. 
In general, $\varphi$ decreases with increasing $B/T$ and $C_{28}$,
with Pearson correlation coefficients of $\rho$\,=\,$-0.28$ and $\rho$\,=\,$-0.46$, respectively, although the
scatter is substantial.
These results are at odds with the conclusions
of \cite{Hart2017}, who reported a slight tendency for $\varphi$ to rise with 
increasing ratio of bulge mass to total mass, precisely the {\it opposite}\ of 
what we see.  Contrary to previous studies \citep{Seigar1998, Kendall2015}, 
we find a significant correlation between $\varphi$ and $C_{28}$: 
$\varphi$ decreases from $=23\fdg7\pm4\fdg8$ at $C_{28}$\,=\,$2.0\pm0.2$ to 
$13\fdg4\pm6\fdg1$ at $C_{28}$\,=\,$5.0\pm0.2$.  The marked scatter in the $\varphi 
- B/T$ diagram may stem, in part, from the many complications in 
bulge-to-disk decomposition \citep{Gao2017}.
Our results imply that galaxies with more centrally concentrated mass 
distributions tend to have more tightly wound spiral arms.

\subsection{Dependence on Luminosity and Mass}

\begin{figure*}
\figurenum{5}
\centering
\includegraphics[width=16cm]{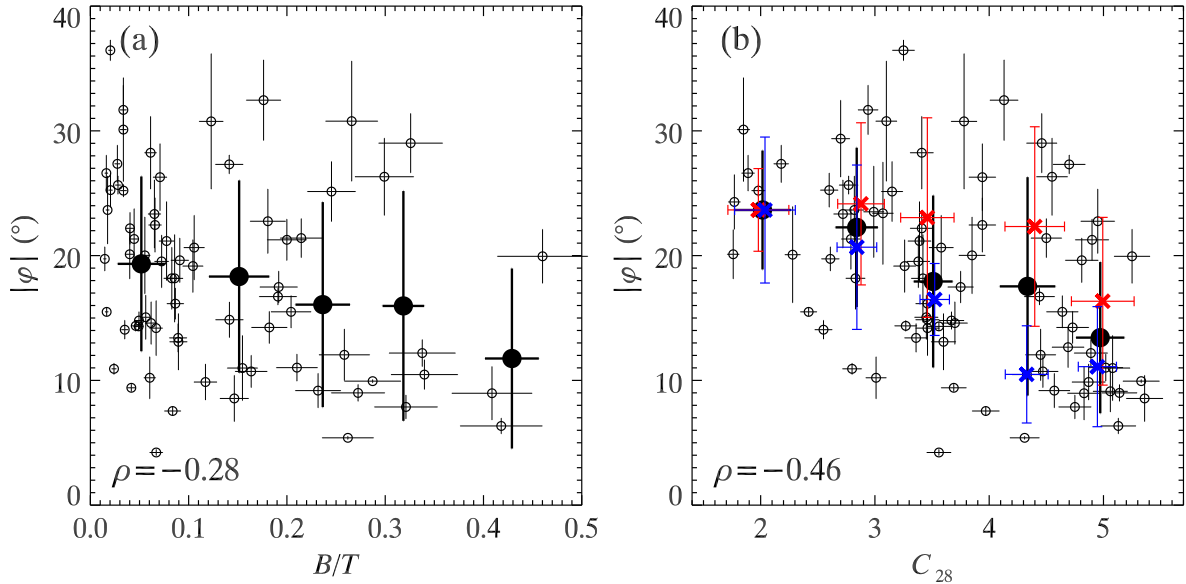}
\caption{
Variation of pitch angle of spiral arms with (a) bulge-to-total light ratio $B/T$ and (b) concentration 
index $C_{28}$. 
The data points are separated into five equal-sized bins of $B/T$ or $C_{28}$.
The mean value and standard 
deviation of each bin is marked by solid black points and associated error 
bars.  Pitch angle correlates weakly with $B/T$ and somewhat stronger with 
$C_{28}$.  For each of the five bins of $C_{28}$, the data are further 
separated into two subsets according to the mean value of $M_*^{\rm gal}$; 
the mean and standard deviation of the subset above and below the mean are 
marked by the blue and red crosses, respectively.
The Pearson's correlation coefficient, $\rho$, is shown on the bottom-left of each panel.}
\end{figure*}
\begin{figure*}
\figurenum{6}
\centering
\includegraphics[width=15cm]{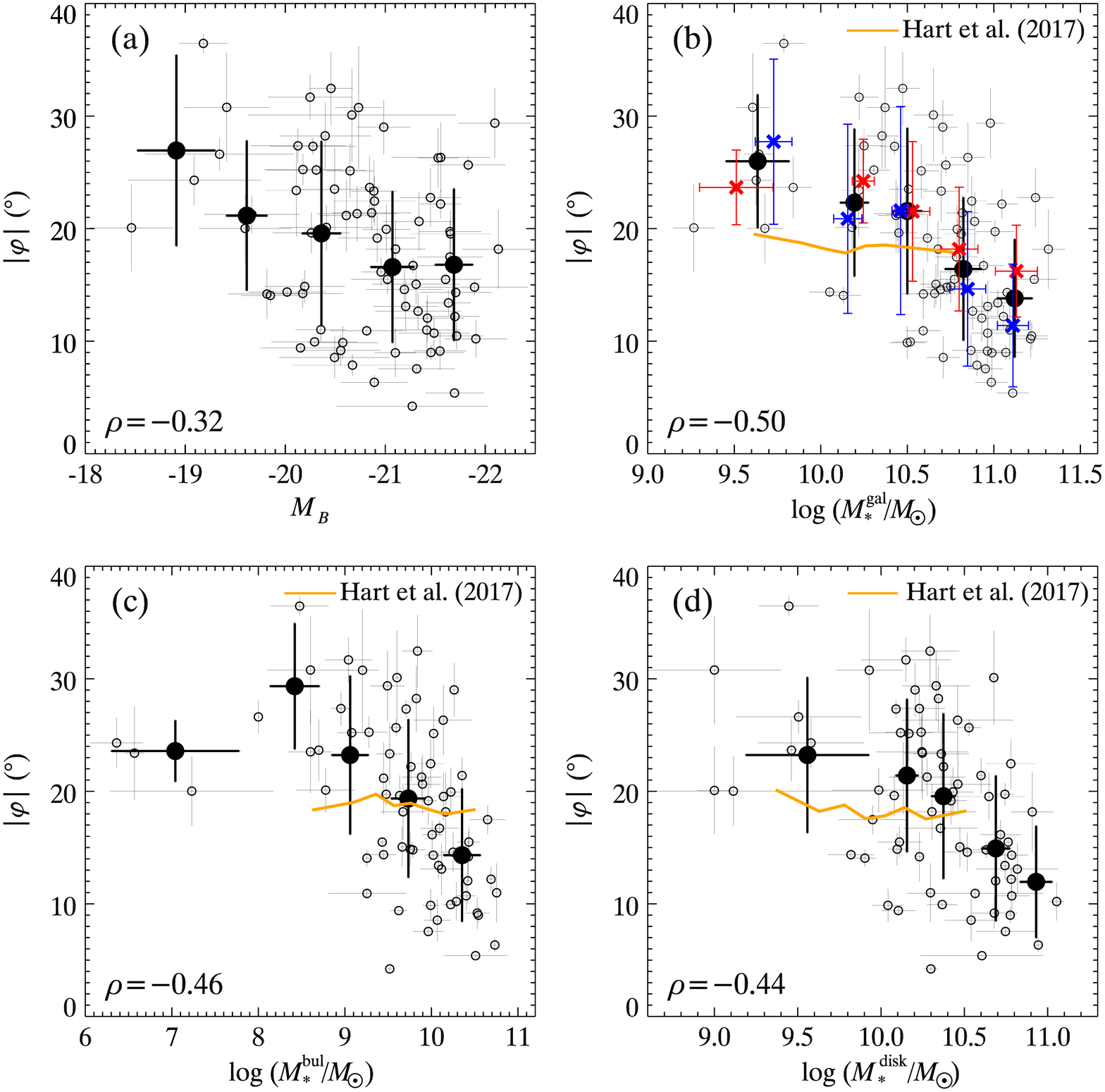}
\caption{Dependence of pitch angle on (a) absolute {\it B}-band magnitude 
($M_B$), (b) galaxy stellar mass ($M_*^{\rm gal}$), (c) bulge stellar mass 
($M_*^{\rm bul}$), and (d) disk stellar mass ($M_*^{\rm disk}$). 
The solid black points and their associated error bars mark the mean value and 
standard deviation of the data in each of the five bins of the parameter on the
X-axis value.  
To obtain sufficient data points, the number of objects in the first bin was 
manually adjusted to include all sources up to
$M_B = -19.3$, $M_*^{\rm gal} = 10^{10}$\,$M_{\odot}$, $M_*^{\rm bul} = 10^8$\,$M_{\odot}$, and $M_*^{\rm disk} = 10^{10}$\,$M_{\odot}$, and 
the rest of the data were further grouped into four equal bins.
Results from \cite{Hart2017} are marked by orange lines,
which show a nearly flat trend.  For each of the five bins of $M_*^{\rm gal}$,
the data are further separated into two subsets according to the mean value of 
$C_{28}$; the mean and standard deviation of the subset above and below 
the mean are marked by the blue and red crosses, respectively.
The Pearson's correlation coefficient, $\rho$, is shown on the bottom-left of each panel.}
\end{figure*}

Figure~6 examines the variation of pitch angle with absolute {\it B}-band 
magnitude ($M_B$), total galaxy stellar mass ($M_*^{\rm gal}$), and 
separately the stellar mass of the bulge ($M_*^{\rm bul}$) and disk 
($M_*^{\rm disk}$).  The results from \cite{Hart2017}, marked by the 
orange line, are shown for comparison.  
As shown in Figure~6, the distributions of the data are not homogenous, and
there are fewer data points in the faint, low-mass end. 
Thus, we manually adjust the number of objects in the first bin 
to include all sources up to
$M_B = -19.3$, $M_*^{\rm gal} = 10^{10}$\,$M_{\odot}$, $M_*^{\rm bul} = 10^8$\,$M_{\odot}$, and $M_*^{\rm disk} = 10^{10}$\,$M_{\odot}$, and
the rest of the data were further grouped into four equal-sized bins.
The binned data support the notion 
that more luminous, more massive galaxies tend to have more tightly wound 
spiral arms (smaller values of $\varphi$).
The apparent bimodal distribution at $M_B\approx -20.7$ mag is probably an 
artifact
of the small number of data points in the range from 8$\degr$ to 32$\degr$.
The correlation between $\varphi$ and $M_B$ is mainly driven by the close 
coupling between $M_B$ and $M_*^{\rm gal}$.  
The measured pitch angles 
decrease from $\varphi$\,=\,$26\fdg0\pm5\fdg9$ at $\log (M_*^{\rm gal}/M_{\odot})$\,=\,$9.6\pm0.2$ to $\varphi$\,=\,$13\fdg8\pm5\fdg3$ at $\log (M_*^{\rm gal}/M_{\odot})$\,=\,$11.1\pm0.1$.  
Similarly, pitch angles decrease toward larger $M_*^{\rm bul}$ and $M_*^{\rm disk}$, 
with scatter comparable to that of the $\varphi-M_*^{\rm gal}$ relation.  The 
apparent flattening or even reversal toward the lowest mass bin is probably 
caused by insufficient sampling.  As illustrated by the orange lines, 
\cite{Hart2017} find essentially no relationship between pitch angle and 
stellar mass (total, bulge, or disk), and at a given mass their pitch angles 
are systematically smaller than ours.  The discrepancy is likely caused by the 
usage of different techniques to measure pitch angle.  The Fourier 
transformation we employ uses flux as weighting when calculating Fourier 
spectra, and so it tends to extract information from the dominant modes of 
the spiral structure.  \cite{Hart2017}, by contrast, employ the code 
{\tt SpArcFiRe} \citep{sparcfire2014}, which uses texture analysis to identify 
arc-like segments, including very faint arms, and then averages the pitch 
angles of these segments, weighting by their length.  However, the faint arms, 
whose physical significance is unclear, may adversely affect their final 
measurement of pitch angle.

Our findings are qualitatively consistent with the theoretical expectations of 
density wave theory.  
\cite{Roberts1975} argued that the mass concentration determines the pitch
angle of spiral arms, a conclusion confirmed by the modal analysis of 
\cite{Bertin1989}.  In a similar vein, \cite{Hozumi2003} suggested that 
tighter spiral arms are associated with higher surface density.  In Figure~6b, 
we study the effect of mass 
concentration on arm tightness at {\it fixed}\ galaxy mass, by grouping the 
data into bins of $\log (M_*^{\rm gal})$ after dividing the sample into two 
subsets according to their mean value of  $C_{28}$.  For a given stellar mass 
$\log (M_*^{\rm gal}/M_{\odot})\gtrsim10.4$, we note that galaxies with 
higher $C_{28}$ (blue crosses) tend to have smaller $\varphi$, but the 
tendency disappears for stellar masses $\log (M_*^{\rm gal}/M_{\odot})
\lesssim10.4$, likely due to small number statistics.  At the same time, at 
fixed $C_{28}$ more massive galaxies tend to have tighter arms; the difference 
decreases with decreasing $C_{28}$ and vanishes at $C_{28}\approx 2$ 
(Figure~5b).

\subsection{Dependence on Galaxy Kinematics}

\subsubsection{Central velocity dispersion}

\begin{figure}
\figurenum{7}
\centering
\includegraphics[width=8cm]{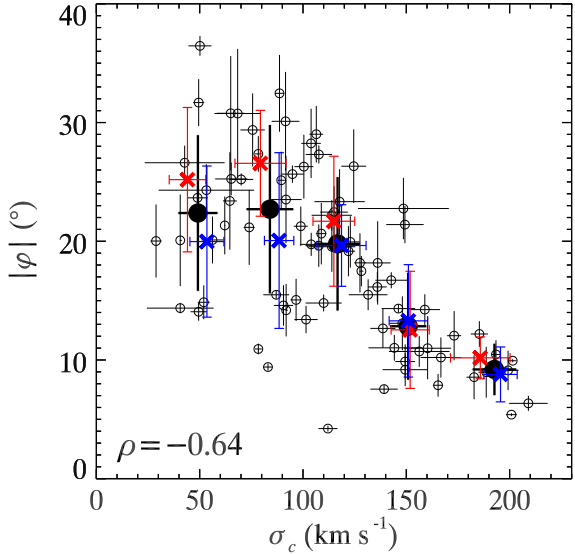}
\caption{Correlation between pitch angle and central velocity dispersion ($\sigma_c$).
The data points are grouped into five equal bins of $\sigma_c$.
The solid black points and their associated error bars mark the mean value and 
standard deviation of the data in each bin.
The pitch angle decreases with increasing $\sigma_c$, with small scatter 
for $\sigma_c\ga100$\,km\,s$^{-1}$, but for $\sigma_c\lesssim100$\,km\,s$^{-1}$ 
the mean value of pitch angle remains roughly constant.  For the the five bins 
of $\sigma_c$, the data points are further separated into two parts according 
to the mean value of $\log(M_*^{\rm gal})$; the mean and standard deviation of 
the top and bottom parts are marked by the blue and red crosses, respectively.
The Pearson's correlation coefficient, $\rho$, is shown on the bottom-left corner.
}
\end{figure}

Figure~7 plots pitch angles versus the central velocity dispersion 
($\sigma_c$).  The correlation is strong, with
with Pearson correlation coefficient $\rho$\,=\,$-0.64$.
For galaxies with $\sigma_c\ga100$\,km\,s$^{-1}$, pitch angle 
decreases with $\sigma_c$ with small scatter, reaching a mean value of 
$\varphi = 9\fdg2\degr\pm2\fdg2$ at $\sigma_c=193\pm11$\,km\,s$^{-1}$, the 
highest velocity dispersion covered by our sample.  No galaxies with high 
$\sigma_c$ show open spiral arms.  By contrast, galaxies with 
$\sigma_c\lesssim100$\,km\,s$^{-1}$ host arms with a wide spread in pitch 
angles, from values as high as $\varphi \approx 30\degr$ to as low as
$\varphi \approx 15\degr$.  The pitch angle seems to remain at a roughly 
constant mean value of $\varphi \approx 23\degr$ for 
$\sigma_c\lesssim100$\,km\,s$^{-1}$.

\subsubsection{Comparison with previous results}

\cite{Seigar2008} reported a strong inverse correlation between spiral arm 
pitch angle central stellar velocity dispersion for a sample of 27 galaxies.
Here, we independently reexamine their results.  Instead of using the same 
images used in \cite{Seigar2008}, we use, whenever possible, images of higher 
quality: seven images from CGS \citep{Ho2011}, 16 images from SDSS, and 
three images collected from the NASA/IPAC Extragalactic Database 
(NED)\footnote{\tt http://nedwww.ipac.caltech.edu}. The data point for the 
Milky Way is not included, as its pitch angle may be unreliable.  We uniformly 
analyze all the objects, using, for consistency, the 2D Fourier transformation 
method.  \cite{Seigar2008} did not provide the projection parameters ($e$ and 
PA) for the galaxies, and so measure them following the procedure described in 
Section 3.1.  We successfully measure pitch angles for 21 of the 26 galaxies. 

Direct comparison of our new pitch angle measurements with those published by 
\cite{Seigar2008} reveal that a significant fraction of them (6/21) were 
severely and systematically overstimated (by more than $8\degr$).  Figure~8 
illustrates the four cases with the most serious discrepancy, whose pitch 
angles were overestimated by more than $10\degr$.  The panels in left column 
present the unsharp-masked images overplotted with the synthetic arm with the 
pitch angle derived from this work (red solid line) and from the work of 
\citeauthor{Seigar2008} (\citeyear{Seigar2008}; orange dash-dotted line). The panels in the right column
plot the 2D Fourier spectra, with an arrow indicating the peak chosen to 
derive the pitch angle.  It is obvious that the synthetic arms created using 
our pitch angles trace the spiral arms very well, whereas those that adopt the 
pitch angles from \cite{Seigar2008} do not.  In the extreme case of NGC~3938 
(top row), \cite{Seigar2008} quoted a pitch angle of $43\fdg4$, whereas we 
find $16\fdg7$. 
These four galaxies have clear spiral arms and a distinctly dominant Fourier 
mode.  Their pitch angles can be measured rather straightforwardly and 
unambiguously. The large discrepancies with the published values cannot arise 
from errors in the measurement technique.

With our updated pitch angle measurements in hand for the 21 galaxies 
reanalyzed by us, we redraw in Figure~9 the relation between $\varphi$ and 
$\sigma_c$ as originally published by \cite{Seigar2008}, using $\sigma_c$ 
given by those authors (their Table~1).  The relationship between $\varphi$ 
and $\sigma_c$ is considerably less tight than claimed by \cite{Seigar2008}, 
and instead closely resembles our results based on a much larger sample 
(Figure~7).

\begin{figure*}
\figurenum{8}
\centering
\includegraphics[width=12cm]{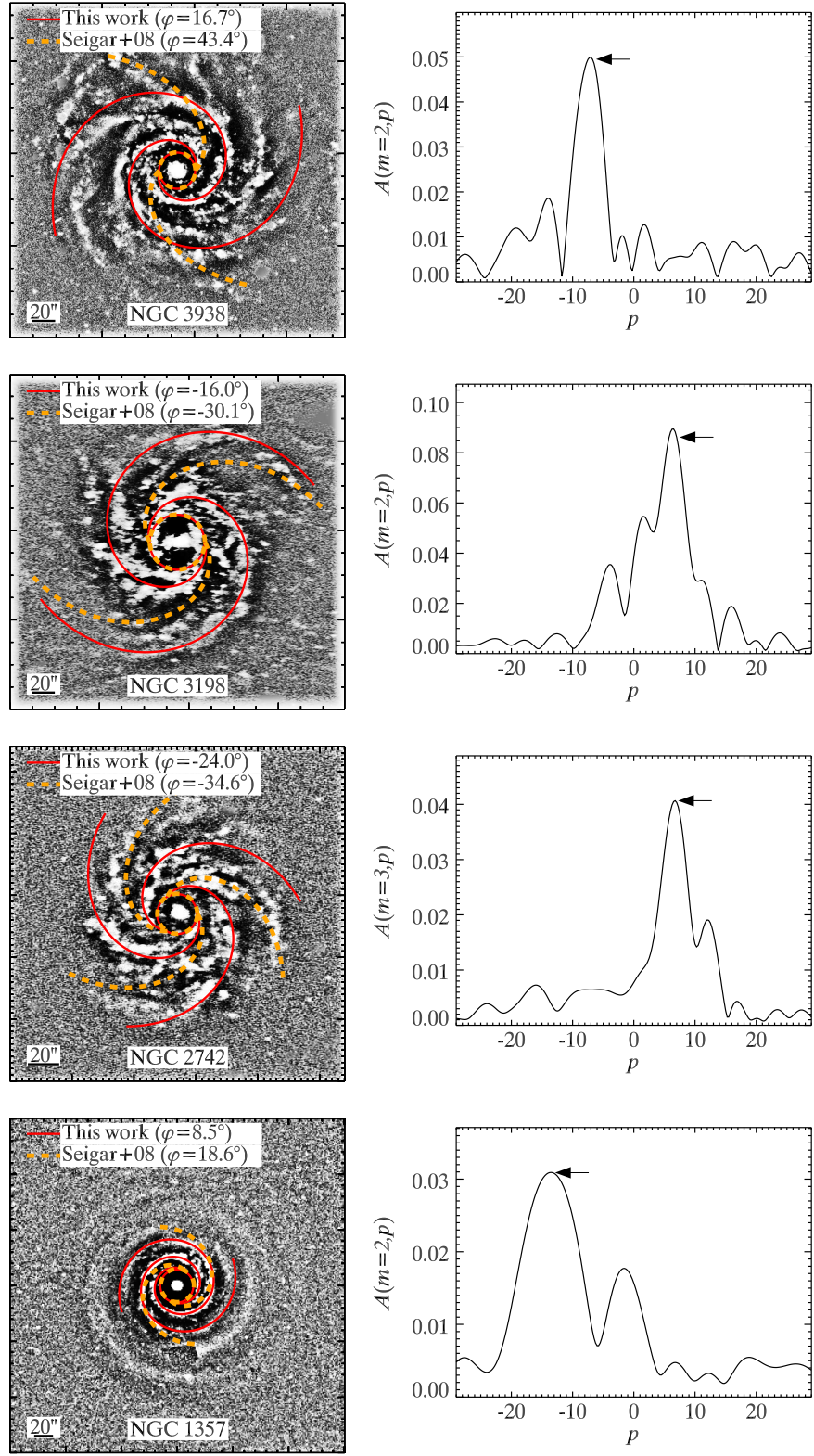}
\caption{
Illustration of our new measurements of pitch angle for four galaxies in common with the study of \cite{Seigar2008}.  (Left) Unsharp-masked image overplotted with synthetic arms with our pitch angle 
(red-solid curve) and that of \citeauthor{Seigar2008} (\citeyear{Seigar2008}; dashed-orange curve). (Right) Fourier spectra 
to derive the pitch angle, with an arrow indicating the peak chosen.
}
\end{figure*}

\begin{figure}
\figurenum{9}
\centering
\includegraphics[width=8cm]{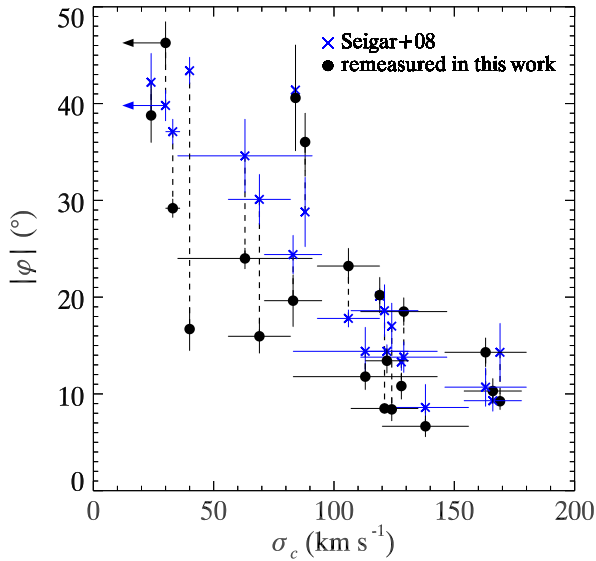} 
\caption{Correlation between pitch angle and $\sigma_c$. Blue crosses mark the 
results from \cite{Seigar2008}, while black solid points represent our new 
measurements. Data points for the same galaxy are connected with a dashed line 
for comparison. Our new measurements show that there are objects with pitch 
angles $\sim20\degr$ at $\sigma_c\approx 50$\,km\,s$^{-1}$, making the 
trend very similar to that presented in Figure~7.
}
\end{figure}

\subsubsection{Implications}

\begin{figure}
\figurenum{10}
\centering
\includegraphics[width=8cm]{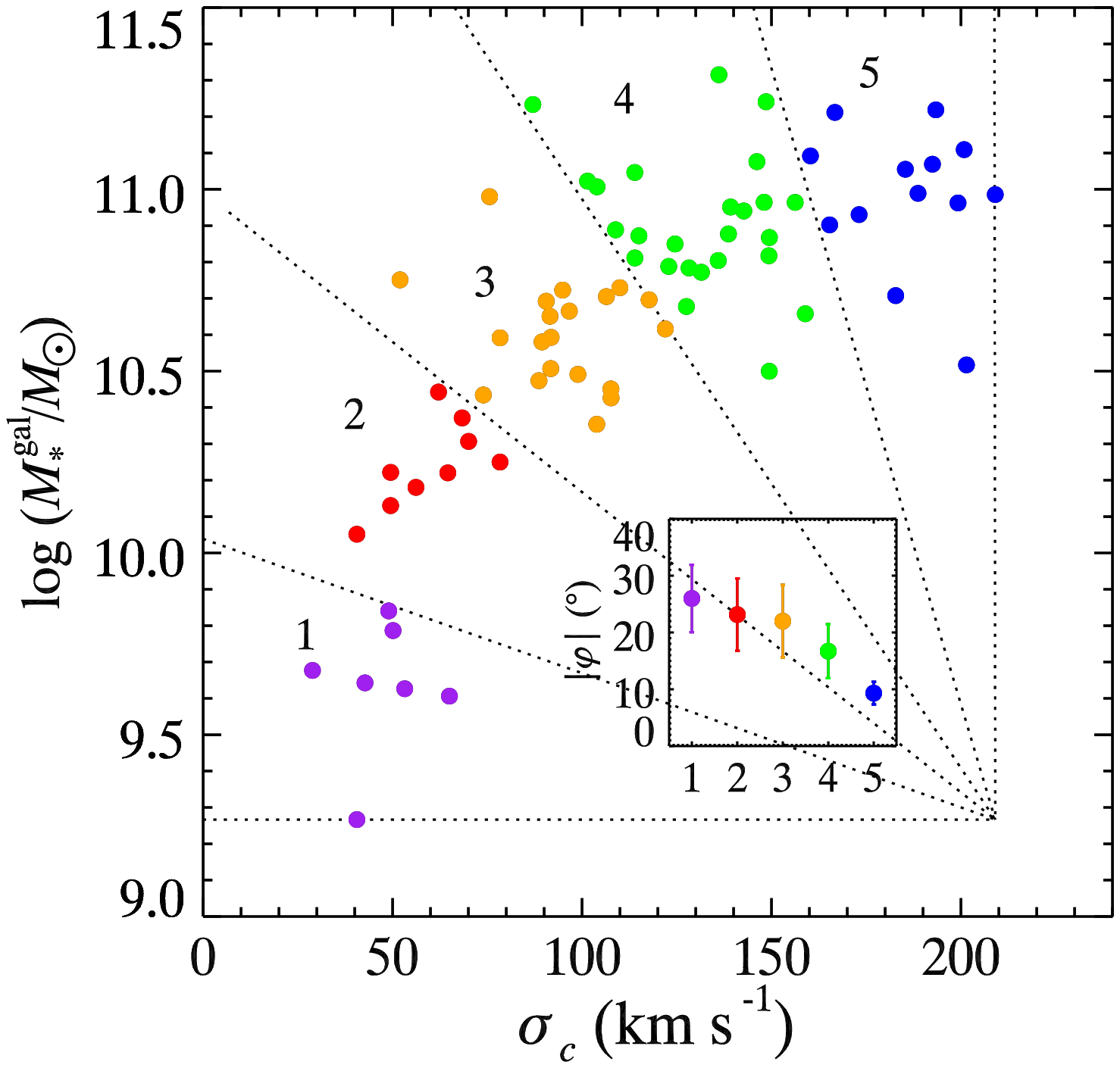}
\caption{Galaxy stellar mass $(M_*^{\rm gal})$ is plotted against central 
velocity dispersion ($\sigma_c$).  The data are further separated into five 
fan-shaped bins, denoted by number of 1 (purple), 2 (red), 3 (orange), 4 
(green), and 5 (blue). The mean value and standard deviation of pitch angle in 
each bin are shown in the inset panel: (1) $26\fdg0\pm6\fdg0$, (2) 
$23\fdg1\pm6\fdg4$, (3) $21\fdg9\pm6\fdg4$, (4) $16\fdg7\pm4\fdg7$, and (5) 
$9\fdg3\pm2\fdg0$. Our results suggest that when $\sigma_c$ or 
$M_*^{\rm gal}$ is high, the pitch angle is mainly determined by 
$\sigma_c$, while when $\sigma_c$ or $M_*^{\rm gal}$ is low, the pitch 
angle is mainly determined by $M_*^{\rm gal}$.
}
\end{figure}

Pitch angle correlates strongly with $\sigma_c$ for 
$\sigma\ga100$\,km\,s$^{-1}$.  As $\sigma_c$ is measured within the central 
$3\arcsec$ ($\sim 1$ kpc for our median sample distance of 68 Mpc), it must 
connect with certain global properties to influence galactic-scale spiral 
structure.  It is well known that the luminosity of elliptical galaxies scales 
with stellar velocity dispersion following $L\propto\sigma^4$ \citep{Faber1976},
but the Faber-Jackson relation of bulges is not well-determined.  It varies 
systematically with Hubble type \citep{Whitmore1981, Kormendy1983} and between 
classical bulges and pseudobulges \citep{Kormendy2004}.  The correlation 
between pitch angle and $\sigma_c$ is not entirely consistent with that 
between pitch angle and bulge mass.  The $\varphi - M_*^{\rm bul}$ relation 
clearly has larger scatter than the $\varphi - \sigma_c$ relation, especially 
at the high-mass end (Figure~6c).  For $\sigma_c\lesssim100$\,km\,s$^{-1}$, 
the mean value of pitch angle remains essentially constant, suggesting that in
this regime another parameter determines the pitch angle.  In order to study 
the effect of galaxy mass on pitch angle for a given central velocity 
dispersion, for each of the five bins of $\sigma_c$, we again split the data 
into two subsets according to the mean value of $\log(M_*^{\rm gal})$, and then
separately examine the behavior of each.  As Figure~7 shows, when 
$\sigma_c\ga100$\,km\,s$^{-1}$ stellar mass does not help to reduce the 
scatter in pitch angle, but in the low-$\sigma_c$ regime more massive galaxies 
tend to have smaller pitch angle, consistent with the empirical relation found 
in Section 4.2.  Thus, galaxy mass can account for part of the scatter in 
pitch angle.

The $\varphi-M_*^{\rm gal}$ relation and the $\varphi-\sigma_c$ 
relation are actually projections of a stronger three-parameter relation 
involving $\varphi$, $M_*^{\rm gal}$, and $\sigma_c$. Figure~10 plots
$\log(M_*^{\rm gal})$ against $\sigma_c$. 
To separate the data equally into five fan-shaped bins, numerically denoted 1 
to 5, we scale the data by dividing them by 
($\sigma_{c, {\rm max}}$\,$-$\,$\sigma_{c, {\rm min}}$) and 
($\log M^{\rm gal}_{*, {\rm max}}$\,$-$\,$\log M^{\rm gal}_{*, {\rm min}}$) 
and then generate six radiant dotted lines 
in the scaled parameter space, with equal separation in orientation angle of 
$18\degr$,  originating from a reference point with maximum $\sigma_c$ and 
minimum $M_*^{\rm gal}$. 
The counterparts of the dotted lines in the original ($\sigma_c$, $\log M^{\rm gal}_*$) space are shown in Figure~10.
We calculate the mean and standard deviation of $\varphi$ in each bin.  
The inset panel in Figure~10 shows the tendency 
of pitch angle progressively increasing from the fifth bin 
($\varphi$\,=\,$9\fdg3\pm2\fdg0$) to the first bin 
($\varphi$\,=\,$26\fdg0\pm6\fdg0$).  Our results suggest that when $\sigma_c$ 
or $M_*^{\rm gal}$ is high, $\varphi$ is mainly determined by $\sigma_c$, 
whereas when $\sigma_c$ or $M_*^{\rm gal}$ is low, $\varphi$ is mainly 
determined by $M_*^{\rm gal}$.  Figure~10 can explain the behavior of the 
$\varphi-M_*^{\rm gal}$ and $\varphi-\sigma_c$ relations.  In the high-mass 
regime, the scatter in $\varphi$ is large at a given $M_*^{\rm gal}$ (Figure 
5b) because $\sigma_c$ is high ($\sim 130-210$ km\,s$^{-1}$) and $\varphi$ is 
mainly dictated by $\sigma_c$.  In the low-$\sigma_c$ regime, the mean value of 
$\varphi$ remains nearly constant with $\sigma_c$ (Figure~7) because 
$\varphi$ is mainly determined by $M_*^{\rm gal}$ instead of $\sigma_c$. 
Therefore, our results suggest that two primary parameters---central velocity 
dispersion and galaxy mass---synergistically determine spiral arm pitch angle.

\subsubsection{Morphology of the Rotation Curve of the Central Region}

\begin{figure}
\figurenum{11}
\centering
\includegraphics[width=8cm]{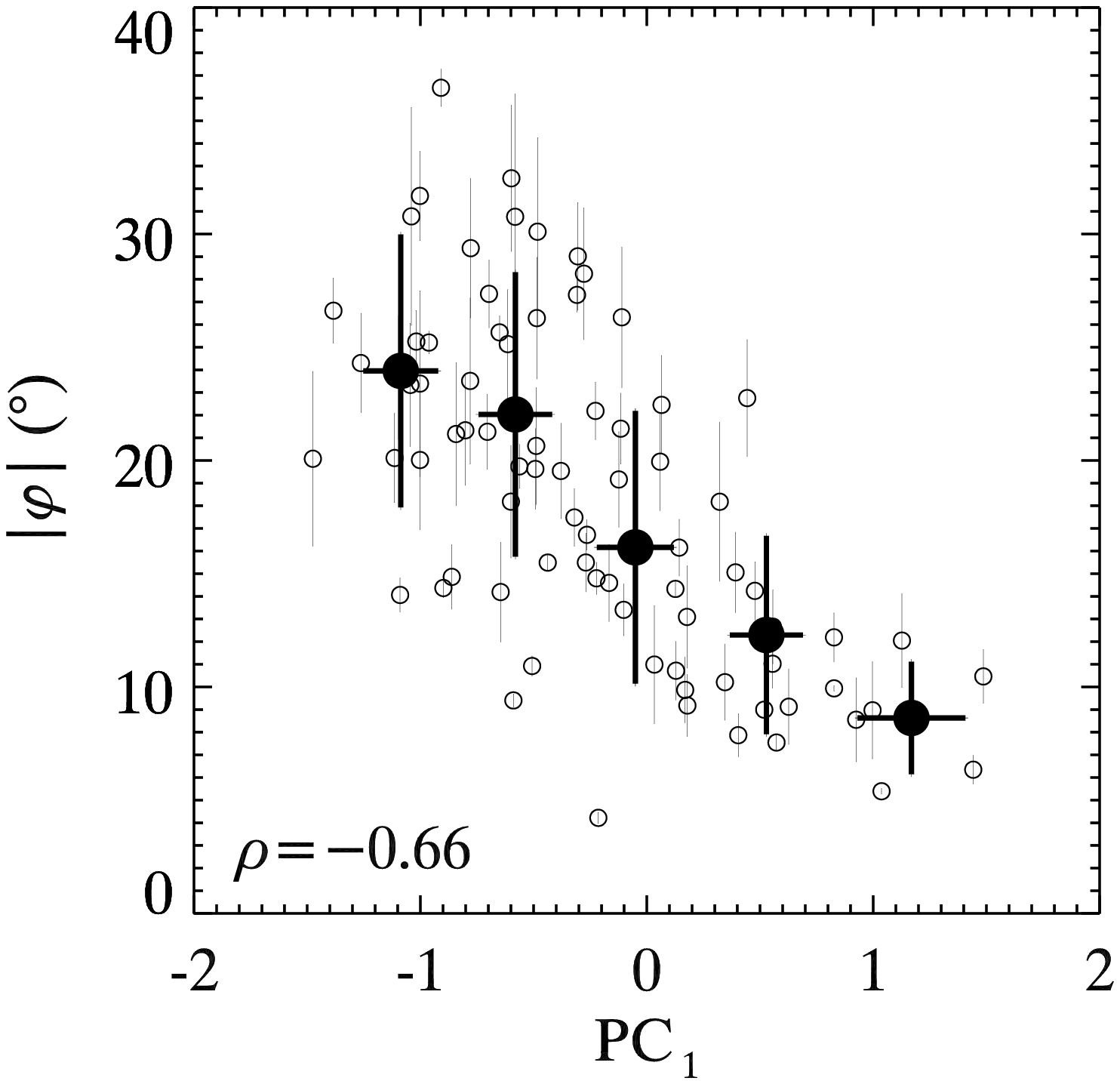}
\caption{The pitch angle shows a strong inverse correlation with PC$_1$. 
The data points are grouped into five equal bins of PC$_1$.
The solid black points and their associated error bars mark the mean value and 
standard deviation of the data in each bin.  PC$_1$ 
reflects the morphology of the rotation curve in the central region.  Galaxies 
with high PC$_1$ ($>0$) have a high-amplitude, centrally peaked rotation curve 
that attains a sharp maximum near the center, followed by a dip and then a 
broad maximum of the disk component; those with low PC$_1$ ($<0$) have a 
low-amplitude, slow-rising rotation curve that rises gently from the center in 
a rigid-body fashion.
The Pearson's correlation coefficient, $\rho$, is shown on the bottom-left corner. 
}
\end{figure}

The observed rotation curves can be grouped roughly qualitatively into several 
types according to their behavior in the central region 
\citep[e.g.,][]{Keel1993, Sofue1999}.  \cite{Kalinova2017} applied principal 
component analysis to quantitatively classify the rotation curves of the 
CALIFA galaxies. The coefficient of the first eigenvector 
PC$_1$ describes the morphology of the rotation curve of the central region.
Galaxies with high PC$_1$ ($>0$) have a high-amplitude, centrally peaked 
rotation curve that attains a sharp maximum near the center, followed by a 
dip and then a broad maximum of the disk component; those with low PC$_1$ 
($<0$) have a low-amplitude, slow-rising rotation curve that rises gradually 
from the center in a rigid-body fashion.  As the coefficient PC$_1$ is a 
measure of both the shape and amplitude of the rotation curve, PC$_1$ 
simultaneously reflects the mass of the central component and the mass of 
the disk, especially for bulgeless galaxies. As expected from the 
$\varphi\,-\,\sigma_c\,-\,M_*^{\rm gal}$ relation,  $\varphi$ shows a 
strong inverse correlation with PC$_1$ (Figure~11; 
$\rho$\,=\,$-0.66$).

In addition to baryonic mass, the shape and the amplitude of the rotation 
curve also reflect the mass distribution of dark matter in galaxies.  Hence, 
the strong correlation between pitch angle
and morphology of the central rotation curve also implies that dark matter 
content might help to shape spiral arms.

\subsubsection{Shear rate}

\begin{figure}
\figurenum{12}
\centering
\includegraphics[width=8cm]{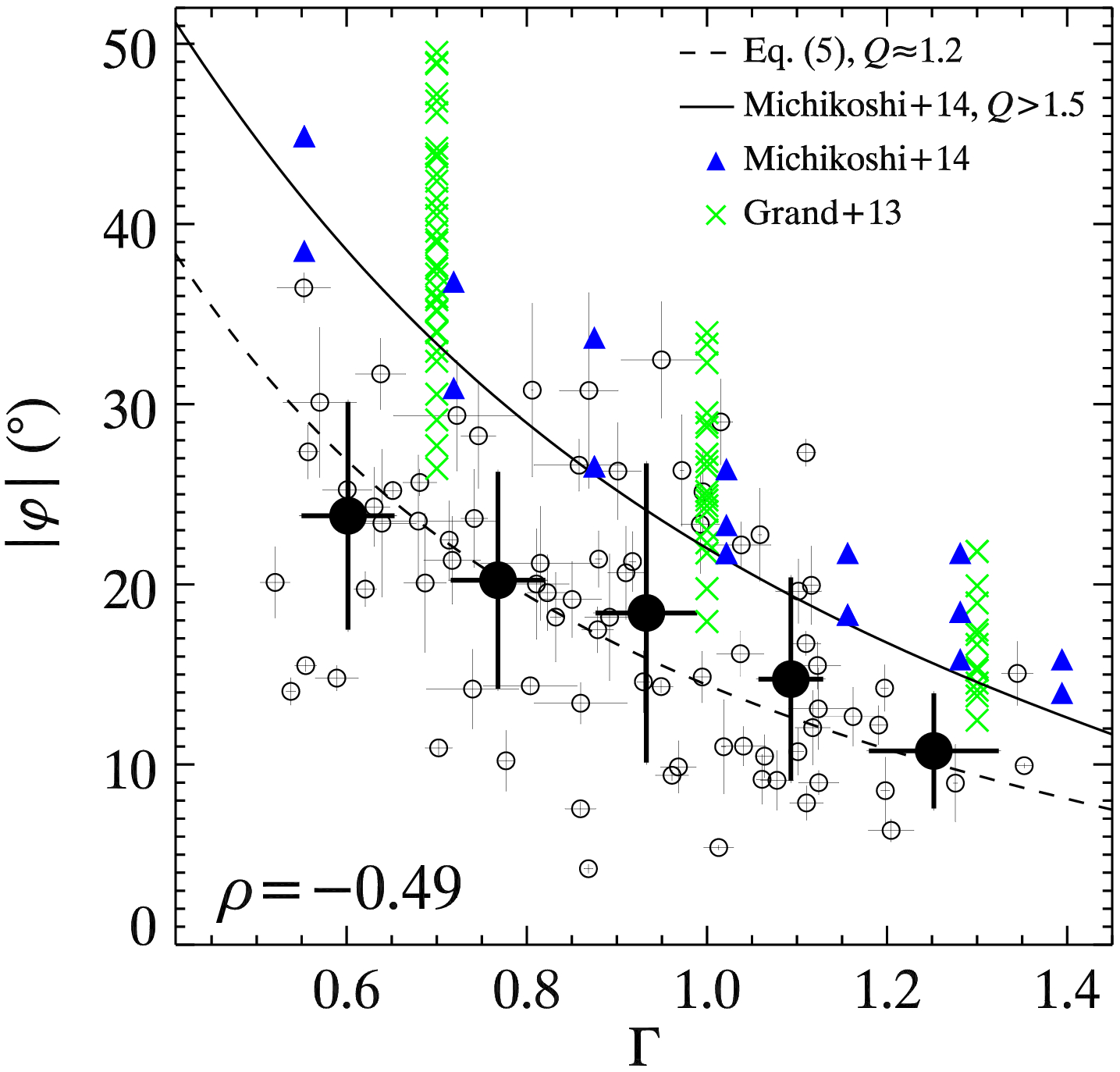}
\caption{Comparison between pitch angle and shear rate.  
Our measurements are 
plotted as small open black points, with large solid black points denoting the 
mean and associated standard deviation for five equal bins in $\Gamma$. The results 
from $N$-body simulations of \cite{Grand2013} and \cite{Michikoshi2014} are 
plotted as green crosses and blue triangles, respectively. The black solid 
curve traces the theoretical prediction of swing amplification theory, given 
by \cite{Michikoshi2014}, assuming $Q>1.5$. The dashed curve denotes the 
prediction for $Q\approx 1.2$ (Eq. (5)).
The Pearson's correlation coefficient, $\rho$, is shown on the bottom-left corner. 
}
\end{figure}

The strong correlation between pitch angle and shear rate $\Gamma$ originally 
suggested by \citep{Seigar2005, Seigar2006} has not been substantiated by the 
recent study of \cite{Yu2018a}, who show that $\sim 1/3$ of the pitch angle 
measurements of \cite{Seigar2006} have been severely overestimated, as a 
consequence of which the correlation between $\varphi$ and $\Gamma$ is much 
weaker than previously claimed\footnote{It behooves us to note that we 
uncovered similar problems with the pitch angle measurements in 
\cite{Seigar2005}. For example, these authors report $\varphi = 30\fdg4\pm1.9$
for ESO~576$-$G51, but the Fourier spectra shown in their Figure~2 clearly 
demonstrate that its dominant $m=2$ mode reaches its peak at $p\approx 6$, 
which corresponds to only $\varphi \approx 18\degr$.  Inspection of the Fourier
spectra of other galaxies in their study (e.g., ESO~474$-$G33) reveals similar 
inconsistencies.}.  We reassess the relationship between $\varphi$ and $\Gamma$
in Figure~12.
Pitch angle does have a tendency to decline with increasing
$\Gamma$ ($\rho$\,=\,$-0.49$), although the scatter is large. 
Two physical mechanisms may explain this behavior.

An association between $\varphi$ and $\Gamma$ was recently explored using 
numerical simulations by \cite{Grand2013} and \cite{Michikoshi2014}.  $N$-body 
simulations of isolated stellar disks produce transient but recurrent 
local spiral arms \citep{Carlberg1985, Bottema2003, Sellwood2011, Fujii2011,
Grand2012a, Grand2012b, Baba2013, Onghia2013}.  \cite{Onghia2013} argue that 
spiral arms originate from the dynamical response of a self-gravitating 
shearing disk to local density perturbations.  Differential motion tends to 
stretch and break up the spiral arms locally.  In regions where self-gravity 
dominates, the disk is locally overdense and generates arm segments, which 
reconnect and make up the spiral arms.  The simulations of \cite{Grand2013} 
and \cite{Michikoshi2014}, marked by green crosses and blue triangles in 
Figure~12, share the same general tendency for $\varphi$ to decline with 
increasing $\Gamma$.  However, a systematic offset can be clearly seen. 
For a given shear rate, the predicted pitch angles are larger than the observed
values, typically by $\Delta\varphi\ga8\degr$, especially at the low-$\Gamma$ 
end.  This implies that, in terms of arm morphology, $N$-body simulations 
cannot yet generate realistic spiral arms even if the resolution of the 
simulations is very high.

Swing amplification \citep{Julian1966, Goldreich1978, Toomre1981} is a 
mechanism of amplifying spiral arms when a leading spiral pattern rotates 
to a trailing one due to the shear in a differentially rotating disk.
Swing amplification theory is reasonably consistent $N$-body simulations in 
terms of the predicted number of arms, which is approximately inversely 
proportional to the mass fraction of the disk component \citep{Carlberg1985, 
Bottema2003, Fujii2011, Onghia2013}, and in terms of the relationship between 
the shear rate of the rotation curve and the pitch angle of the simulated 
spirals \citep{Grand2013, Michikoshi2014}.  

\cite{Michikoshi2014} derived a theoretical relation between $\varphi$ and 
$\Gamma$ in the context of swing amplification theory.  We give a brief 
summary here. We consider a material arm that swings from leading to trailing 
due to differential motion.  The pitch angle evolves as
\citep[e.g.,][]{Binney}

\begin{eqnarray}
\tan\varphi=\frac{1}{2At},
\end{eqnarray}

\noindent
where $A$ is the first Oort constant and $t=0$ means that the arm extends
radially outward across the disk ($\varphi$\,=\,$90\degr$). 
In the simulations of \cite{Michikoshi2014}, Toomre's $Q$ parameter increases
rapidly and exceeds 1.5 for $\Gamma\gtrsim 0.5$.  In the linear approximation 
of swing amplification theory, if $Q>1.5$, the maximum amplification is reached
at $t_{\rm max} \backsimeq3.5/\kappa$, where $\kappa$ is the epicyclic 
frequency.  They interprete the most amplified short wave as the spiral 
structures observed in their simulation.  Combined with 
$\Gamma=\frac{2A}{\Omega}$, the relation between pitch angle and shear rate 
becomes (their Eq. 15)

\begin{eqnarray}
\tan\varphi = \frac{2}{7} \frac{\sqrt{4-2\Gamma}}{\Gamma}.
\end{eqnarray}

\noindent
This is indicated by the solid line in Figure~12.  The prediction from swing 
amplification theory is quantitatively consistent with the results from 
$N$-body simulations.  Note that the predicted pitch angle for a given shear 
rate is still higher than our measurements (by $\sim8\degr$).  This is because 
Eq. (4) assumes $Q>1.5$, which is indeed the case for the simulated spiral 
galaxies of \cite{Michikoshi2014}.  However, as suggested by \cite{Bertin1989},
because of self-regulation of gas content, Toomre's $Q$ parameter in the outer 
regions of galactic disks should be close to unity.  On the other hand, $Q$ may
be substantially larger in the central regions of galaxies because of the 
presence of a bulge.  Considering both the effects of the disk and the bulge, 
if $Q$ is constrained to $\sim 1.2$, we estimate from Figure 7 of 
\cite{Michikoshi2014} that $t_{\rm max} \approx 5.5/\kappa$.  For 
$Q \approx 1.2$, we obtain, from swing amplification theory, 

\begin{eqnarray}
\tan\varphi = \frac{2}{11} \frac{\sqrt{4-2\Gamma}}{\Gamma}.
\end{eqnarray}

\noindent
This revised relation, presented as the dashed curve in Figure~12, is now 
consistent with our observational results.  
Our results have two important implications.  First, if swing amplification 
theory is the correct framework to explain spiral structures in galaxies, then 
Toomre's $Q$ should be roughly 1.2.  And second, to generate more 
realistic spiral arms, simulations may need an effective cooling mechanism for 
the disk, perhaps by including the effects of gas, to lower Toomre's $Q$.

In the framework in which spiral structure constitute transient material arms, 
the shape of the arms should reflect the effects of differential rotation alone.
Although we present some evidence supporting this picture, the existence of 
other stronger empirical relationships 
between pitch angle and $\sigma_c$ or PC$_1$, which have statistically 
stronger correlation coefficients,
probably rules out the transient material arms scenario.

A more compelling, alternative explanation for the relationship between pitch 
angle and shear rate comes from density wave theory.  As the shape of the 
rotation curve depends on the distribution of mass, the shear rate reflects
the mass concentration. Consequently, the inverse correlation between $\varphi$ 
and $\Gamma$ is qualitatively consistent with the expectations of density wave 
models \citep{LinShu64, Roberts1975, Bertin1989, Bertin1989b}.  
The predicted relation betweem $\varphi$ and $\Gamma$ is consistent with our 
observed $\varphi-C_{28}$ relation (Figure~5b).

\section{Summary and Conclusions}

After more than half a century of research, the physical origin of spiral arms 
in galaxies is still a topic of debate \cite[for a review, see][]{Dobbs2014}. 
The pioneering work of \cite{Kennicutt1981} systematically established the 
dependence of spiral arm pitch angle ($\varphi$) on galaxy properties, but 
there have been only a handful of quantitative follow-up studies since \citep{Ma2002,
Seigar2005,Seigar2006,Seigar2008,Kendall2011,Kendall2015,Davis2015,Hart2017}.  
The CALIFA survey of nearby galaxies provides a good opportunity to revisit 
this problem, given the plethora of relevant ancillary measurements available 
for the sample, including luminosity, stellar mass, photometric decomposition, 
and kinematics \citep{Walcher2014,Sanchez2016Pipe3D,FBJ2017,Sanchez2017,JMA2017,
Catalan-Torrecilla2017, Gilhuly2018}.  Because of the low redshift of the 
sample, SDSS images are adequate to resolve the spiral structure of the 
galaxies \citep{Yu2018a}.  This paper uses SDSS $r$-band images to perform a 
detailed analysis of the spiral structure of the 79 relatively face-on CALIFA 
spiral galaxies with available published rotation curves.  We aim to 
systematically examine the correlation between spiral arm pitch angle and 
other galaxy properties to investigate the physical origin of spiral arms. 

As implicit in the Hubble classification system \citep{Hubble1926, Sandage1961},
we confirm that spiral arms become more open from early to late-type galaxies.
The pitch angle of spiral arms decreases with brighter absolute magnitude, 
larger stellar mass (total, bulge, or disk), and higher concentration 
($C_{28}$), although all these correlations contain significant scatter. Pitch 
angle is also correlated with $B/T$.  For a given $M_*^{\rm gal}$, galaxies 
with higher $C_{28}$ have tighter spiral arms.  Similarly, for a given $C_{28}$,
more massive galaxies have tighter spirals. 
These trends are consistent with the density wave theory for spirals, which
predicts that pitch angle decreases with higher mass concentration 
\citep{Roberts1975} and larger surface density \citep{Hozumi2003}.
We also find a strong correlation between pitch angle and 
central stellar velocity dispersion: 
$\sigma_c\ga100$\,km\,s$^{-1}$, $\varphi$ decreases with increasing $\sigma_c$ 
with small scatter, whereas $\varphi$ remains roughly constant for 
$\sigma_c\lesssim100$\,km\,s$^{-1}$.  This bevavior has important 
implications.  We show that $\varphi$ is mainly determined by $\sigma_c$ for 
massive galaxies, while the primary determinant of $\varphi$ becomes 
$M_*^{\rm gal}$ for less massive galaxies.  We then demonstrate that the 
$\varphi - M_*^{\rm gal}$ and $\varphi - \sigma_c$ relations are projections
of higher dimensional relationship between $\varphi$, $M_*^{\rm gal}$, and 
$\sigma_c$.

Spiral arm pitch angle is closely connected to the morphology of the central
rotation curve, quantified by PC$_1$, the coefficient of the first eigenvector 
from principal component analysis.  Galaxies with centrally peaked rotation 
curves tend to have tight arms; those with slow-rising rotation curves tend to 
have loose arms.  As PC$_1$ reflects both the mass of the central component 
and of the disk, especially for bulgeless galaxies, the connection between 
pitch angle and the morphology of the central rotation curve is consistent 
with the $\varphi - \sigma_c - M_*^{\rm gal}$ relation.

We do not confirm the strong connection between pitch angle and galactic 
shear rate ($\Gamma$) found in previous studies.  $N$-body simulations 
\cite[e.g.,][]{Grand2013, Michikoshi2014}, while generally successful in 
reproducing the qualitative dependence between $\varphi$ and $\Gamma$, 
systematically overpredicts $\varphi$ at fixed $\Gamma$.  The observed inverse
correlation between $\varphi$ and $\Gamma$ can be interpreted in the context 
of transient material arms obeying swing amplification theory, provided that
Toomre's $Q \approx 1.2$.  In this scenario, the shape of the spiral arms 
reflects the effects of differential rotation alone.  

Differential rotation, however, is likely not the primary determinant of spiral
arm pitch angle, as the empirical correlation between $\varphi$ and $\Gamma$ is
not as strong as those between $\varphi$ and $\sigma_c$ or PC$_1$.  The 
totality of the evidence places greater weight on the density wave theory for 
the origin of spiral arms in galaxies.

\begin{acknowledgements}
This work was supported by the National Key R\&D Program of China 
(2016YFA0400702) and the National Science Foundation of China (11473002, 
11721303).  We are grateful to an anonymous referee for helpful feedback.
\end{acknowledgements}

\end{document}